\documentclass[12pt]{iopart}

\usepackage{iopams}
\usepackage{graphicx}
\usepackage[section]{placeins} 
\usepackage{wrapfig}
\usepackage{subfig} 
\begin{document}

\title[Non-symmorphic band degeneracy at the Fermi level in ZrSiTe]{Non-symmorphic band degeneracy at the Fermi level in ZrSiTe}

\author{Andreas Topp$^1$, Judith M. Lippmann$^{1,2}$, Andrei Varykhalov$^3$, Viola Duppel$^1$, Bettina V. Lotsch$^{1,2,4}$, Christian R. Ast$^1$ and Leslie M. Schoop$^1$}
\address{$^1$ Max Planck Institute for Solid State Research, Heisenbergstr. 1, 70569 Stuttgart, Germany}
\address{$^2$ Department of Chemistry, Ludwig-Maximilians-Universit\"at M\"unchen, Butenandtstr. 5-13, 81377 M\"unchen, Germany}
\address{$^3$ Helmholtz-Zentrum Berlin f\"ur Materialien und Energie, Elektronenspeicherring BESSY II, Albert-Einstein-Stra\ss e 15, 12489 Berlin, Germany}
\address{$^4$ Nanosystems Initiative Munich (NIM) \& Center for Nanoscience, Schellingstr. 4, 80799 M\"unchen, Germany}
\eads{\mailto{a.topp@fkf.mpg.de}~~ \mailto{c.ast@fkf.mpg.de}~~ \mailto{l.schoop@fkf.mpg.de}}

\vspace{10pt}
\begin{indented}
\item[]\today
\end{indented}

\begin{abstract}
Non-symmorphic materials have recently been predicted to exhibit many different exotic features in their electronic structures. These originate from forced band degeneracies caused by the non-symmorphic symmetry, which not only creates the possibility to realize Dirac semimetals, but also recently resulted in the prediction of novel quasiparticles beyond the usual Dirac, Weyl or Majorana fermions, which can only exist in the solid state. Experimental realization of non-symmorphic materials that have the Fermi level located at the degenerate point is difficult, however, due to the requirement of an odd band filling. In order to investigate the effect of forced band degeneracies on the transport behavior, a material that has such a degeneracy at or close to the Fermi level is desired. Here, we show with angular resolved photoemission experiments supported by density functional calculations, that ZrSiTe hosts several fourfold degenerate Dirac crossings at the X point, resulting from non-symmorphic symmetry. These crossings form a Dirac line node along XR, which is located almost directly at the Fermi level and shows almost no dispersion in energy.  ZrSiTe is thus the first real material that allows for transport measurements investigating Dirac fermions that originate from non-symmorphic symmetry.
\end{abstract}

%
\vspace{2pc}

\submitto{\NJP}
%
\maketitle
%
%

\section{Introduction}

Non-symmorphic space groups force band degeneracies at high symmetry points in the Brillouin Zone (BZ), which makes materials in these space groups ideal candidates for Dirac semimetals (DSMs) in two or three dimensions \cite{young2012dirac,young2015dirac}. In contrast to ``conventional'' Dirac materials (e.\ g.\ graphene), such specially protected band crossings cannot be lifted by spin orbit coupling (SOC) allowing for two dimensional (2D) DSMs without a gap \cite{young2015dirac}. Besides fourfold degeneracies, which result in normal Dirac cones, non-symmorphic space groups can also force three-, six-, or eight fold degeneracies which has recently lead to the prediction of new quasiparticles, beyond Dirac, Weyl or Majorana, in non-symmorphic materials \cite{bradlyn2016beyond,wieder2016double}. \\

The problem in realizing such non-symmorphic DSMs is their requirement of an odd band filling to locate the Fermi level at the degenerate point. More specifically, to have a ``clean'' non-symmorphic DSM, an isolated half-filled band is required \cite{gibson2015three}. This can be understood if one considers that the translational part of the non-symmorphic symmetry causes a folding of the BZ, which results in the forced band degeneracies. Hence, these degeneracies are located at half filling of the band. However, non-symmorphic degeneracies can be at the Fermi level also in charge balanced systems, if they are stabilized by other pockets. It should be noted that here we are counting electrons per formula unit rather than per unit cell, as commonly done for non-symmorphic systems (See SI for further explanation) \cite{watanabe2016filling,watanabe2015filling}. Half-filled bands are very difficult to obtain in real materials, since compounds with such band fillings are commonly unstable, and typically undergo Peierls transitions \cite{hoffmann1987chemistry,dixit1984electron} or become Mott insulators \cite{mott1990metal}. For example, compounds with $d^1$ filling, such as TaS$_2$ or TaSe$_2$, are well-known for forming charge density waves and undergoing Peierls transitions \cite{wilson1975charge}. Compounds with other partial $d$-fillings typically undergo Mott transitions and/or localize antiferromagnetically to become insulating. Typical examples are transition metal oxides such as MnO, CoO, NiO \cite{austin1970metallic,zaanen1985band}, or the cuprates \cite{lee2006doping}. Isolated, half-filled \textit{s} or \textit{p} bands are even more difficult to stabilize. A famous example is BaBiO$_3$ with a formally half-filled \textit{6s} band. However in reality, in BaBiO$_3$ the unit cell is doubled with two distinct bismuth sites where the \textit{6s} band on one site is filled and on the other site it is empty \cite{baumert1995barium}. Electron-doped BaBiO$_3$ has also recently been proposed to be a topological insulator \cite{yan2013large}. The same type of charge order also appears in many other compounds with formally half-filled \textit{s} bands, examples being CsTlCl$_3$ \cite{schoop2013lone,retuerto2013synthesis}, Pb$_3$O$_4$ \cite{terpstra1997electronic} or Sn$_2$S$_3$ \cite{mootz1967kristallstruktur}. Another common way to avoid half-filling for $p$-block materials is to form polyanionic Zintl compounds \cite{corbett1985polyatomic}. Here, metal-metal bonds cause localization of the electrons and thus semiconducting behavior. Other materials that formally have half-filled bands and do not localize or distort are typically stabilized by another band that overlaps with the half-filled band. Examples are the alkali metals or coppermonochalcogenides (e.g. CuS) \cite{zaanen1985band}.
For these reasons, there is currently no example of a real material that has a non-symmorphic Dirac crossing located at the Fermi level. In order to study the effect of band degeneracies forced by non-symmorphic symmetry on the transport behavior, it is crucial to find a material where such a band crossing is located at the Fermi level. Here we present an alternative route to achieve a material with a non-symmorphic band crossing at the Fermi level, opening the possibility to study the transport behavior of these exotic Dirac fermions.\\

ZrSiS, a material that recently received attention for its Dirac line node at the Fermi level \cite{schoop2016dirac,ali2016butterfly,hu2016evidence,neupane2016observation,singha2016titanic,wang2016evidence,takane2016dirac,xu2015two,lv2016extremely,hu2016evidence2,lou2016emergence}, was the first material that had been experimentally shown to exhibit a Dirac cone protected by non-symmorphic symmetry \cite{schoop2016dirac}. This fourfold degenerate point is located about 0.5\,eV below the Fermi level.  ZrSiS is a charge balanced system, so that a non-symmorphic crossing is expected to be above and/or below the Fermi level. Therefore, it is a good starting point to realize such a crossing at the Fermi level. Doping could be one way to achieve this, but doping not only introduces many defects and disorder, making transport experiments more obscure; it also can be chemically challenging. A more promising way to alter the electronic structure is anisotropic strain that changes the hopping integrals between orbitals and, therefore, can shift features of the electronic structure to different energies. Note, that due to ZrSiS's electronic structure, both ways would yield a material where the non-symmorphic degeneracy is stabilized by additional pockets at the Fermi level, hence avoiding an unstable, isolated, half-filled band \cite{schoop2016dirac}. \\

ZrSiS could be a promising material for realizing 2D Dirac fermions protected by non-symmoprhic symmetry, since it shows resemblance to the prediction of such fermions in a hypothetical 2D square net of atoms \cite{young2015dirac,schoop2016dirac}. It has been shown with transport studies, however, that ZrSiS has a rather 3D Fermi surface \cite{ali2016butterfly,hu2016evidence2,singha2016titanic,lv2016extremely,wang2016evidence}. In contrast, analysis of de Haas-van Alphen oscillations revealed that the Fermi surface of the isostructural and isoelectronic compound ZrSiTe exhibits more 2D character \cite{hu2016evidence}.\\

Here, we show with angular resolved photoemission (ARPES) experiments and density functional theory (DFT) calculations that ZrSiTe exhibits a Dirac crossing protected by non-symmorphic symmetry very close to the Fermi level that is stabilized by other pockets located at other parts in the BZ, allowing the non-symmorphic degeneracy to be at the Fermi level in a closed-shell system. Since ZrSiTe is isoelectronic to ZrSiS, this is an example of a charge balanced system with a forced degeneracy at the Fermi level that is stabilized by other pockets. This is a result of the much larger $c/a$-ratio in ZrSiTe as compared to ZrSiS \cite{klein1964zirconium}. Substituting Te for S in ZrSiS is, therefore, equivalent to applying uniaxial tensile strain that shifts the Fermi level, but does not pin it there like a half-filled band would. The fourfold degeneracy at the X point remains due to symmetry retention in ZrSiTe, forming a non-dispersive Dirac line node very close to the Fermi level along the XR line. Further, the interlayer forces of ZrSiTe are very weak, indicating the possibility to create a 2D material with a non-symmorphic Dirac line node at the Fermi level. We report on the first real material that can be studied with transport measurements to investigate Dirac fermions that originate from non-symmorphic symmetry. 

\section{Methods}
\textbf{Sample synthesis and crystal structure characterization}\\
Single crystals of ZrSiTe were synthesized from the constituent elements by mixing stoichiometric amounts of Zr, Si and Te and a small amount of iodine (2~wt\%). The mixtures were sealed in an evacuated quartz tube, heated to 1100$^\circ$C (with a 100$^\circ$C gradient) at a rate of 2$^\circ$C/min and kept at this temperature for three days. Since ZrSiTe is a high temperature phase, a cooling rate of at least 5$^\circ$C/min was applied. Square shaped, plate-like, mm-sized crystals were obtained at the cooler end of the ampoule. All further steps were executed under Ar atmosphere, due to slight air sensitivity of the compound.

Electron microscopy was performed with a Phillips CM30 ST (300 kV, LaB6 cathode). High-resolution transmission electron microscopy (HRTEM) images and precession electron diffraction (PED) patterns were recorded with a complementary metal-oxide-semiconductor (CMOS) camera (TemCam-F216, TVIPS); the microscope is equipped with a nanoMEGAS spinning star to obtain PED images. The program JEMS (Stadelmann) was used to simulate diffraction patterns and HRTEM micrographs. SEM images of crystals were measured with a SEM (Vega TS 5130 MM, Tescan) using a Si/Li detector (Oxford).

\textbf{Electronic structure calculations}
Electronic structure calculations were performed in the framework of density functional theory using the WIEN2K \cite{blaha2001wien2k} code with a full-potential linearized augmented plane-wave and local orbitals [FP-LAPW+lo] basis \cite{singh2006planewaves} together with the Perdew-Burke-Ernzerhof
parametrization of the generalized gradient approximation as the exchange correlation
functional \cite{perdew1996generalized}. The plane-wave cutoff parameter $R_{\rm MT}K_{\rm MAX}$ 
was set to seven and the irreducible Brillouin zone was sampled by 1520 $k$-points (bulk) and by a $30\times 30\times 3$ mesh of k-points (slab). Experimental lattice parameters \cite{bensch1995experimental,klein1964zirconium,wang1995main,onken1964silicid} were used in the calculations. SOC was included as a second variational procedure. The irreducible representations were calculated with the program \textit{irrep}, implemented in WIEN2K. For the slab calculation it was found that cleaving between tellurium atoms resulted in the closest match to the experimental observation in analogy to ZrSiS \cite{schoop2016dirac}. The slab was constructed by stacking 5 unit cells in the \textit{c}-direction that are gapped by a 5.3\,\AA\ vacuum.

\textbf{Angular-resolved photoemission}
For ARPES measurements, crystals were cleaved and measured in ultra-high vacuum (low $10^{-10}$\,mbar range). Low-energy electron diffraction (LEED) measurements showed that the cleavage plane was the (001) plane. ARPES spectra were recorded with the $1^2$-ARPES experiment installed at the UE112-PGM2a beamline at BESSY-II. The measurement temperatures are room temperature (300\,K), 70\,K as well as 40\,K.

\section{Results and Discussion}
\begin{figure}[h]
\centering
\includegraphics[width=1\linewidth]{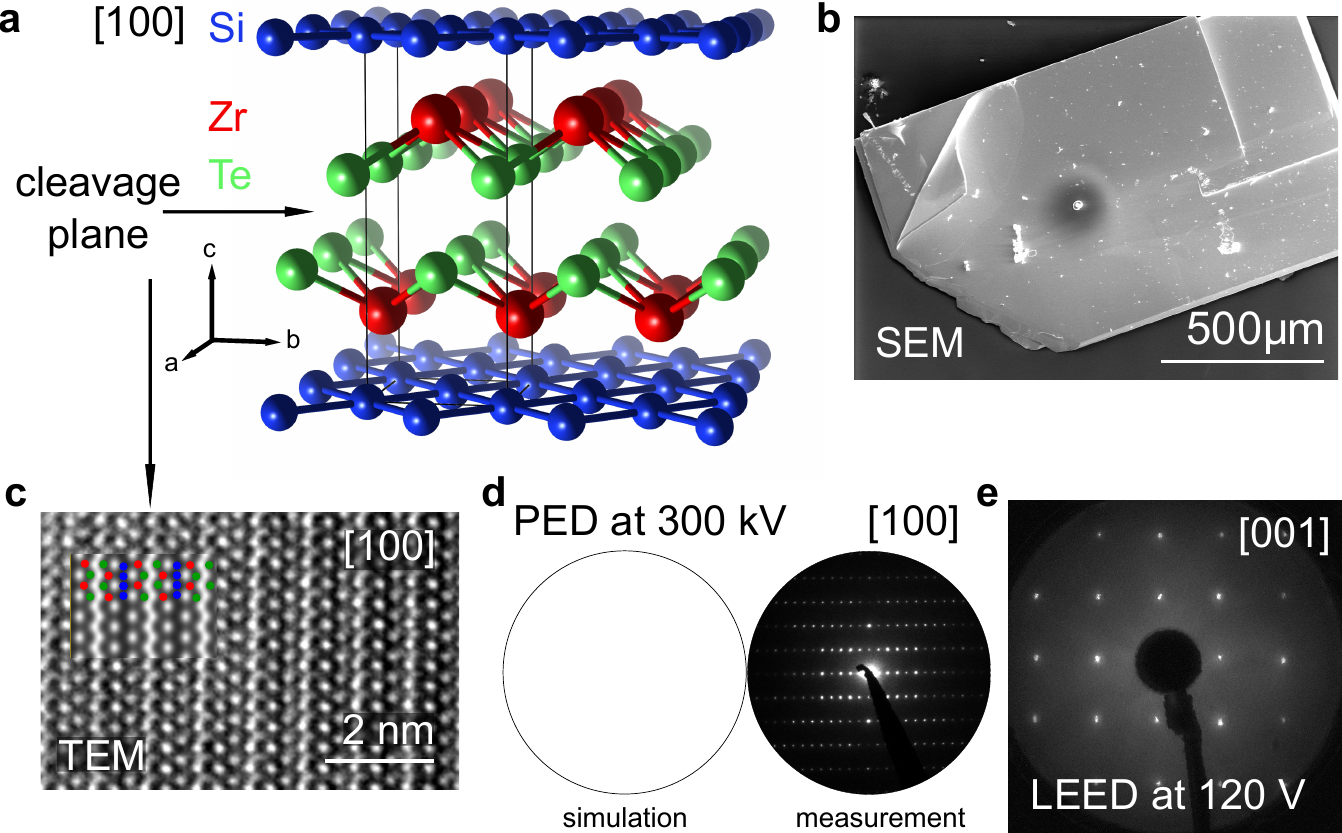}
\captionof{figure}{Crystal structure and characterization of ZrSiTe. \textbf{a} Crystal structure of ZrSiTe in [100] direction with the cleavage plane between the Zr-Te layers. \textbf{b} SEM image of a representative ZrSiTe crystal. \textbf{c} HRTEM micrograph along [100]-direction showing the layered structure and the cleavage plane. The focus plane is $\Delta f = -35\,$nm, with the atoms appearing in black (sample thickness $t=4.79$ nm). The inset shows a simulated image with the atoms marked in corresponding color. \textbf{d} PED pattern of the [100] orientation. Simulated and measured pattern are in excellent agreement. \textbf{e} LEED pattern at 120\,V of the cleaved [001] surface.}
\label{fig:ZrSiTe_structure}
\end{figure}

\textbf{Crystal structure}\\
ZrSiTe, just as the closely related compound ZrSiS, crystallizes in the tetragonal $P4/nmm$ space group (no. 129) adapting the PbFCl structure type. Two double-layers of Zr-Te are separated by a layer of a Si square net. The atomic radius of Te is considerably larger than the one of S leading to an elongation of the lattice constants in ZrSiTe as compared to ZrSiS with a major contribution to the \textit{c}-axis. While the \textit{a} axis increases by only roughly 4\%, the \textit{c} axis increases by about 18\%. The $c/a$-ratio thus increases form 2.27 in ZrSiS to 2.57 in ZrSiTe. The increase in the \textit{c/a} ratio is accompanied by a decrease in the Te-Zr-Te bond angle (78.25$^\circ$ compared to 83.3$^\circ$ in ZrSiS). The crystal structure, displayed in Fig.\ \ref{fig:ZrSiTe_structure}a, was confirmed with powder X-ray diffraction on ground single crystals. Crystals of ZrSiTe, shown in the scanning electron microscopy (SEM) image in Fig.\ \ref{fig:ZrSiTe_structure}b, are rectangular or square shaped due to the tetragonal crystal structure. The crystals can be easily cleaved with adhesive tape, where a thin layer of the crystal is peeled off; a partially peeled off layer is visible in the SEM image. The layered structure of ZrSiTe and the cleavage plane, which is located between the two Zr-Te layers, was imaged with HRTEM in the [100]-direction, as shown in Fig.\ \ref{fig:ZrSiTe_structure}c (a different focus plane can be found in the supplementary Fig.\ 1). Further confirmation of the crystal structure could be gained with electron diffraction techniques. The measured PED pattern of the [100] orientation is in excellent agreement with the simulated pattern shown in Fig.\ \ref{fig:ZrSiTe_structure}d. More PED patterns for different orientations can be found in the supplementary Fig.\ 2. The low energy electron diffraction pattern of the (001) surface, shown in Fig.\ \ref{fig:ZrSiTe_structure}e, shows perfectly square Bragg peaks, confirming the tetragonal lattice.

\begin{figure}[h]
\centering
\includegraphics[width=1\linewidth]{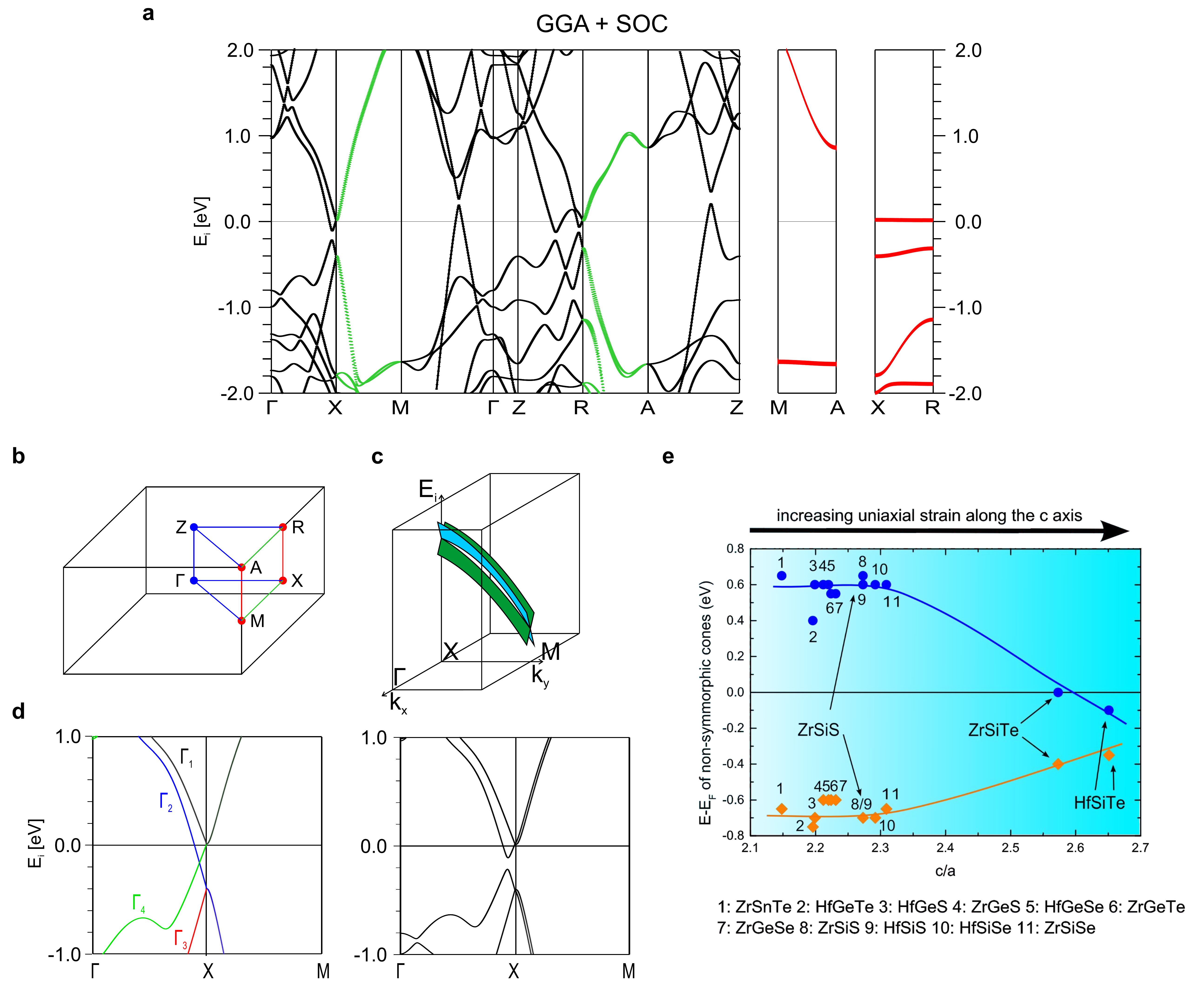}
\captionof{figure}{\textbf{(a)} Calculated bulk band structure with SOC. The degeneracy along the XR and MA direction is marked in red and the XM and RA direction is highlighted in green. \textbf{(b)} BZ of a primitive tetragonal lattice showing the high symmetry points and lines. \textbf{(c)} Schematic drawing of the band crossings that result from non-symmorphic symmetry for $k_z$=0.  \textbf{(d)} Excerpts of the bulk band structure with and without SOC along $\Gamma$XM. While without SOC, there are four different irreducible representations (indicated as $\Gamma_1 - \Gamma_4$), the presence of SOC allows for only one irreducible representation and causes band crossings away from the X point to gap.  \textbf{(e)} A plot showing how location of the Dirac cones (i.e. the fourfold degenerate points) at X (in respect to energy) changes with the $c/a$-ratio for \textit{MXZ} compounds. Blue dots represent the crossing point higher in energy and orange dots represents the one lower in energy. The lines are guides for the eye.}
\label{fig:ZrSiTe_bands_calc}
\end{figure}

\textbf{Bulk band structure}\\
An overview of the calculated bulk band structure of ZrSiTe including SOC can be seen in Fig.\ \ref{fig:ZrSiTe_bands_calc}a. It shows similarities to ZrSiS, as expected from a compound that is isostructral and isoelectronic. As in ZrSiS, the electronic structure features fourfold degeneracies as a result of non-symmorphic symmetry at X, R, A and M, as well as additional Dirac crossings along the $\Gamma$X and $\Gamma$M line. The fourfold degeneracies remain along the XR and MA line (red highlight in Fig.\ \ref{fig:ZrSiTe_bands_calc}a), but are slightly lifted by SOC along the XM and RA line (highlighted in green in Fig.\ \ref{fig:ZrSiTe_bands_calc}a). There are, however, also some striking differences in the electronic structures of the two compounds. The highly dispersed Dirac crossings along $\Gamma$X and $\Gamma$M that form a line node around the Fermi level in ZrSiS, are more gapped in ZrSiTe, which we attribute to the much larger SOC of Te compared to S. The effect of SOC is highlighted in Fig.\ \ref{fig:ZrSiTe_bands_calc}c, showing how the band crossing along the $\Gamma$X line gaps in the presence of SOC, while the degeneracies at X, R, A and M are unaffected due to the non-symmorphic protection. SOC also lifts the degeneracy along the XM and RA line to a larger extent in ZrSiTe than in ZrSiS. Further differences in the electronic structures are unrelated to the magnitude of SOC. In contrast to ZrSiS, where the fourfold degeneracies resulting from non-symmophic symmetry at X are located roughly 0.5\,eV above and below the Fermi level, they are much closer in energy in ZrSiTe (about 400\,meV apart). As a result, one of the cones appears at the Fermi level. In almost all other \textit{MXZ}-phases (\textit{M} = Zr, Hf, \textit{X} = Si, Ge, Sn and \textit{Z} = S, Se, Te), the degeneracies at X are further away from the Fermi level \cite{xu2015two}. ZrSiTe differs from most other \textit{MXZ}-phases in the way that it has a much larger $c/a$-ratio \cite{klein1964zirconium}. Fig.\ \ref{fig:ZrSiTe_bands_calc}d shows how the $c/a$-ratio of the different \textit{MXZ}-phases relates to the energy position of the non-symmorphic Dirac cones (i.e. the fourfold degenerate points). The data has been extracted from the bulk band structures of other known \textit{MXZ} phases, calculated using the published lattice parameters \cite{klein1964zirconium,wang1995main,onken1964silicid}  (see supplemental information Fig.\ 3 and 4 for plots of all bulk band structures). Note that the energy of the cones slightly deviates from measured values due to $k_z$ dependence. While in most phases the $c/a$-ratio has values between 2.2 and 2.3 with the two cones located around 0.6\,eV above and below the Fermi level, the situation is very different for ZrSiTe and HfSiTe (see Fig.\ \ref{fig:ZrSiTe_bands_calc}e). Both compounds have much larger $c/a$-ratios which causes the energy difference between the cones to decrease. In the case of HfSiTe both cones are located below the Fermi level. ZrSiTe, however, has exactly the right $c/a$-ratio to move one cone to the Fermi level. The reason for the much higher $c/a$-ratio can be traced to the different sizes of the Si and Te atoms. While the \textit{a}-axis is related to the Si-Si distance in the Si square net and hence similar to the one of ZrSiS, the Te atoms need more space, which is why the \textit{c}-axis elongates. The resulting uni-axial strain seems to have a strong effect on the electronic structure moving the degeneracies at X towards each other, which results in one crossing being very close to the Fermi level. 

Additionally, it is noteworthy to investigate the dispersion of the bulk band structure perpendicular to the cleaving plane (the $k_z$-direction). In comparison to ZrSiS, ZrSiTe appears much more two dimensional. This observation is not only in line with previous de Haas-van Alphen measurements \cite{hu2016evidence2}, but can also be related to the enlarged $c/a$-ratio of ZrSiTe. Since the Te-Te distance in ZrSiTe is much larger than the S-S distance in ZrSiS, the forces connecting the layers are much weaker. In addition, the electronegativity of Te is lower than the one of S, reducing the ionic bonding character between the layers. This effect can also easily be seen experimentally, since ZrSiTe can be cleaved with adhesive tape while more force is needed to cleave ZrSiS. This enhanced two-dimensionality leads to a very flat dispersion along the XR line in the electronic structure. The R point is located above the X point in the BZ and the four-fold degeneracy is maintained along the line between the two points. Hence, ZrSiTe features a very flat and non-dispersive, four-fold degenerate line node along XR, which is almost directly located at the Fermi level.

\textbf{ARPES measurements}

\begin{figure}[h]
\centering
\includegraphics[width=1\linewidth]{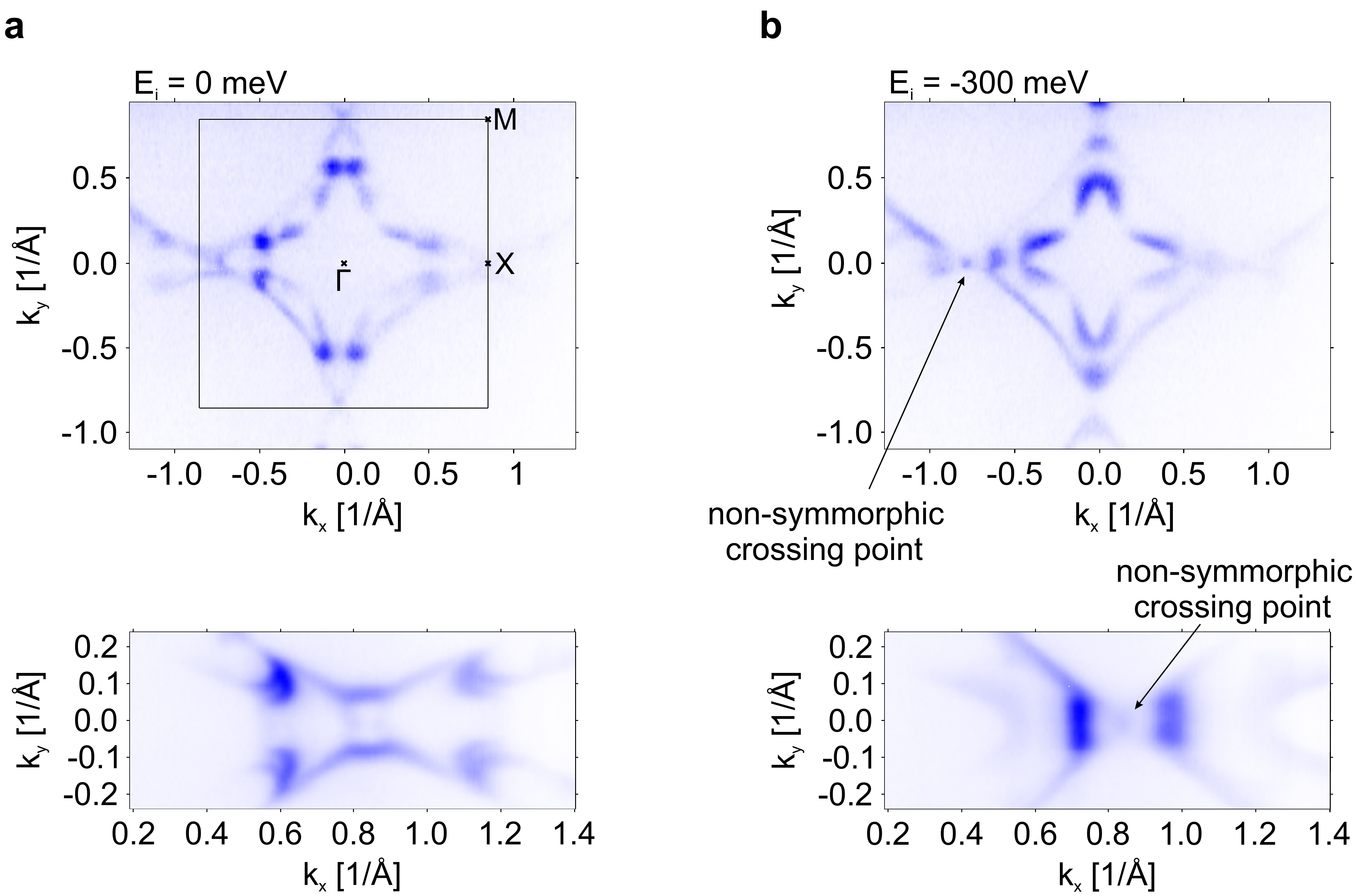}
\captionof{figure}{\textbf{(a)} Constant energy plots at $E_\mathrm{F}$. Top panel shows the whole BZ measured at $h\nu=100\,$eV, while the lower panel shows the surroundings of the X point, measured at $h\nu=30\,$eV. The first BZ is marked by black lines and the high symmetry points $\Gamma$, X and M are labeled. \textbf{(b)} Constant energy plots at $E_\mathrm{F}-300\,$meV. The lower non-symmorphic crossing at X is visible as a dot. Black arrows are indicating the crossing point.}
\label{fig:ZrSiTe_surfaces}
\end{figure}

In order to demonstrate the connection between the $c/a$-ratio and the location of the non-symmophic Dirac cones, we performed ARPES measurements on ZrSiTe. Constant energy surfaces are shown Fig.\ \ref{fig:ZrSiTe_surfaces}. The measured ARPES intensity is plotted against the in-plane wave vectors $k_x$ and $k_y$ for different initial energy cuts at a photon energy of $h\nu=100\,$eV and $h\nu=30\,$eV. The first BZ as well as high symmetry points are highlighted in Fig.\ \ref{fig:ZrSiTe_surfaces}a. The non-symmorphic Dirac cones should appear as point-like features at the X point. To highlight this, we also show magnifications of the area around the X point, measured at $h\nu=30\,$eV. At the Fermi level ($E=0\,$eV), we observe an ellipsoid rather than a point which indicates that the samples are slightly hole doped and the Fermi level is slightly below the non-symmorphic cone. Similarly, crystals of ZrSiS have also been shown to be hole doped \cite{schoop2016dirac,ali2016butterfly}. At an initial energy of about $-300\,$meV the lower non-symmorphic crossing at the X point can be seen as a point (Fig.\ \ref{fig:ZrSiTe_surfaces}b).

\begin{figure}[h]
\centering
\includegraphics[width=1\linewidth]{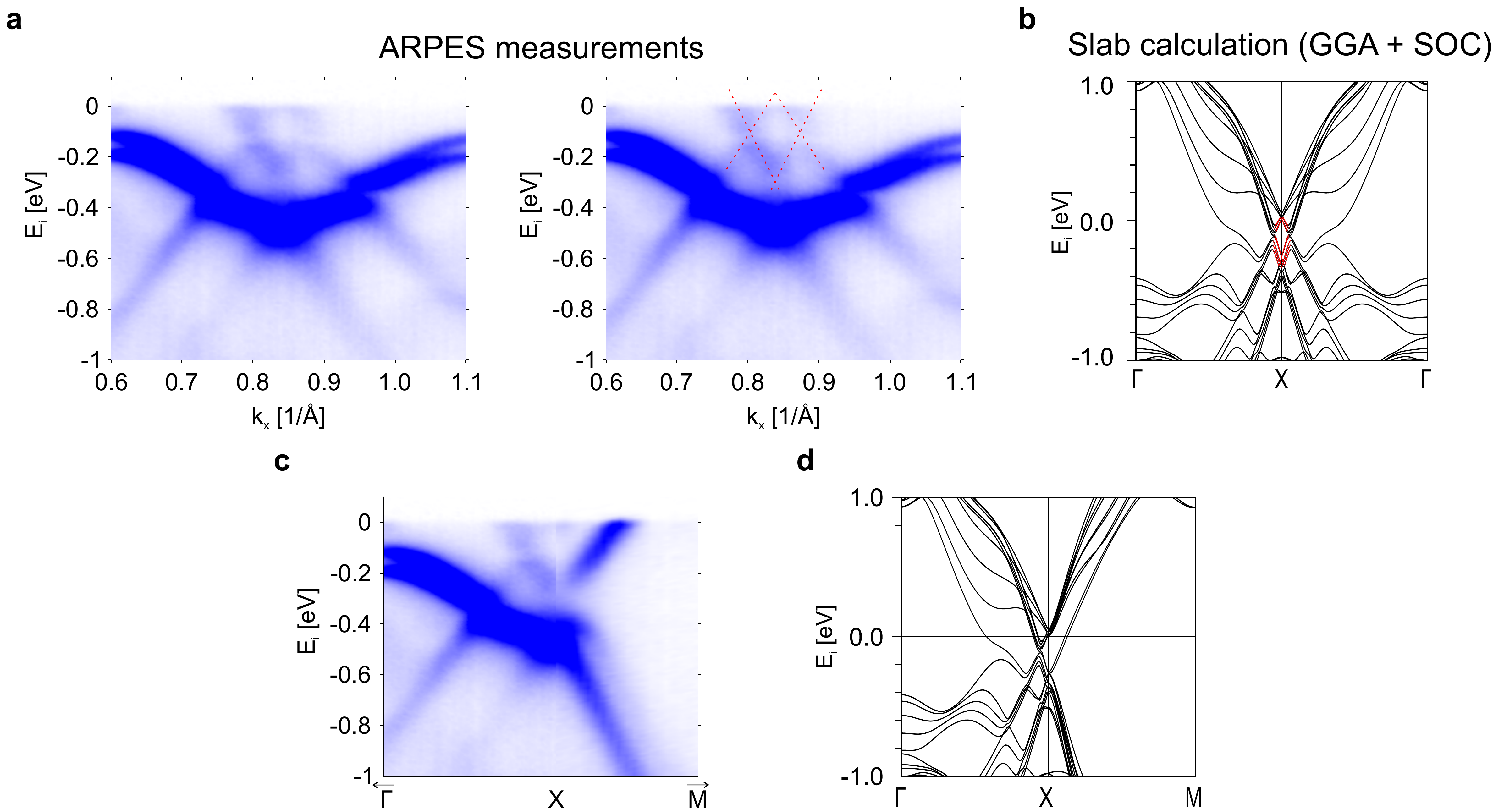}
\captionof{figure}{Dispersion plots along different high-symmetry lines. \textbf{(a)} and \textbf{(b)} show the dispersion along the $\Gamma$X$\Gamma$, \textbf{(c)} and \textbf{(d)} along the $\Gamma$XM direction. The measured ARPES data are shown next to the slab calculations (with SOC). The rhombus-like feature that can be used to identify the degeneracies forced by non-symmorphic symmetry is highlighted in red in both measurement and calculation.}
\label{fig:ZrSiTe_bands}
\end{figure}

Plotting the ARPES data from Fig.\ \ref{fig:ZrSiTe_surfaces} along the high-symmetry lines results in the dispersions shown in Fig.\ \ref{fig:ZrSiTe_bands}. The energy dispersion along the $\Gamma$X$\Gamma$ direction is shown in Fig.\ \ref{fig:ZrSiTe_bands}a. When compared with the bulk band structure shown above, additional states appear in the ARPES data. To test the effect of the surface, we performed calculations of a five unit cell thick slab. The calculations are displayed next to the measured data.
Both, the calculation and ARPES data, show a rhombus-like feature of linear dispersion that is symmetric around the X point at $k_x\approx 0.85\,$\AA$^{-1}$. The upper and the lower corner each represent one of the non-symmophic Dirac cones, while the left and right corner represent the band crossings along the $\Gamma$X line that are gapped by SOC. Dashed lines indicate the rhombus-shaped feature in the measured data. We find the Fermi velocities of these bands to be $\hbar v_\mathrm{F}=3.9$\,eV\AA~ and $\hbar v_\mathrm{F}=5.2$\,eV\AA, which is comparable to ZrSiS \cite{schoop2016dirac}. Note, that in the slab calculation, the gap created by SOC in the rhombus-like feature appears very small due to the dispersion along $k_z$. It is, therefore, not possible to resolve this gap in ARPES. In the measurement, the top cone is slightly above the Fermi level (about $40\,$meV), while the lower cone is at around $-300\,$meV. The observed slight shift in energy of the calculated and measured band structure likely originates from a slightly hole doped sample.  The data prove the presence of a Dirac crossing protected by non-symmorphic symmetry very close to the Fermi level. 

Furthermore, the ARPES data show additional bands that can not be identified as bulk states, but rather as surface states. Similar surface derived states have also been observed in other \textit{MXZ} phases \cite{schoop2016dirac,neupane2016observation,wang2016evidence,takane2016dirac,lou2016emergence}. By overlaying $\Gamma$XM and ZRA (see supplemental information Fig.\ 5), we were able to identify the bands that show surface character, revealing that those are exactly the bands that show high intensity in the ARPES data. (See also Fig.\ 6 in the supplemental information for a surface projection of the calculated band structure. The states originating from the surface Te and Zr atoms mostly contribute to the bands appearing with high intensity in the ARPES measurements.) As can be seen in Fig.\ \ref{fig:ZrSiTe_bands}a,  an intense surface band crosses the lower part of the rhombus. Overall, the agreement between the calculated band structure of a slab and the measured data is very good.
It is easier to identify surface derived states along the XM direction (Fig.\ \ref{fig:ZrSiTe_bands}b), because along this direction there is a gap in the bulk band structure and hence, no bulk bands interfere with the surface derived bands. The upper band that is visible in the ARPES data does not exist in the bulk band structure calculations and is, therefore, surface induced. The surface bands appear to cut the rhombus and then recombine 1\,eV above the Fermi level. This hybridization between surface derived and bulk bands has been observed and discussed in ZrSiS as well and seems to be a typical phenomenon in these types of materials \cite{schoop2016dirac,takane2016dirac}.
Further, if SOC is not included, the bulk bands along the XM line are degenerate, but if SOC is included in the calculation the degeneracy is also lifted, showing that the line along XM is not fully protected by the non-symmorphic symmetry, which is in agreement with the theoretical prediction for square net 2D non-symmorpic materials \cite{young2015dirac}. In contrast to in ZrSiS where this lifting could not be resolved in ARPES it is clearly visible here, due to the higher SOC on Te. A similar effect has been reported for HfSiS, where this effect is even more pronounced \cite{takane2016dirac}.

The ARPES data are generally in very good agreement with the calculated band structure and clearly show that the non-symmorphic Dirac cone is extremely close to the Fermi level. The samples are slightly hole doped, which seems to be common in \textit{MXZ} type phases. Since the compound HfSiTe has a non-symmorphic Dirac cone slightly below the Fermi level in the prediction, it might be located closer to the Fermi level in an experiment due to doping. Therefore, both phases, HfSiTe and ZrSiTe are very promising candidates for investigating the transport behavior originating from bands that are affected by degeneracies forced by non-symmorphic symmetry. Since ZrSiTe can be easily cleaved with adhesive tape and it has already been shown that very thin layers of this compound can be created \cite{klein1964zirconium}, it is also a promising material to create 2D devices of a non-symmorphic Dirac semimetal. To emphasize this, DFT calculations of a monolayer of ZrSiTe were performed showing that the electronic structure of the monolayer is very similar to the bulk compound (see supplemental information Fig.\ 7).

\section{Conclusion}
In summary, we showed that ZrSiTe features a Dirac crossing protected by non-symmorphic symmetry that is located very close to the Fermi level. The fourfold degeneracy at X remains along the XR-line due to the two-dimensionality of the crystal structure. ZrSiTe is therefore the first material where Dirac fermions that are a result of non-symmorphic symmetry can be studied with transport experiments. Since it is difficult to synthesize materials that have half-filled bands, materials with non-symmorphic degeneracies at the Fermi level are extremely rare. We showed that the energy location of the non-symmorphic Dirac cones depends on the $c/a$-ratio of the unit cell and is, therefore, affected by uni-axial strain. Externally applied or intrinsic chemical strain could be used to tune the energy of non-symmorphic degeneracies in other materials as well. For example, strain could be a way to create materials with three-, six-, or eightfold degeneracies at the Fermi level, allowing for transport investigations of new predicted quasiparticles. Thus, we did not only report on the first real material that exhibits a non-symmorphic Dirac crossing at the Fermi level, but also a tool for designing more of such materials.

\ack

The authors thank Raquel Queiroz and Lukas M{\"u}chler for helpful discussions. Financial support by the Max Planck Society, Nanosystems Initiative Munich (NIM) and Center for Nanoscience (CeNS) is gratefully acknowledged. LMS gratefully acknowledges funding from the Minerva fast track fellowship by the Max Planck Society.

\section*{References}

\end{document}